# A Surface-Gated InSb Quantum Well Single Electron Transistor

**J. M. S. Orr[1,2], P.D.Buckle[1], M.Fearn[1], C.J.Storey[1], L.Buckle[1], T.Ashley[1]**
1. QinetiQ. St Andrews Road, Malvern. Worcestershire. WR14 3PS, UK
2. Department of Electrical and Electronic Engineering, University of Manchester, Sackville Street. Manchester. M60 1QW, UK.

**Abstract.** Single electron charging effects in a surface-gated InSb/AlInSb QW structure are reported. This material, due to its large g-factor and light effective mass, offers considerable advantages over more commonly used materials, such as GaAs, for quantum information processing devices. However, differences in material and device technology result in significant processing challenges. Simple Coulomb blockade and quantised confinement models are considered to explain the observation of conductance oscillations in these structures. The charging energy $(e^2/C)$ is found to be comparable with the energy spectrum for single particle states $(\Delta E)$.

**Introduction**

The single electron transistor (SET) relies on the discrete nature of charge to modulate the conductance of a small, isolated volume of conducting material. Within this nanoscale region, known as a quantum dot, the confinement in all three dimensions is sufficiently strong that the electrons may only exist at well defined quantised energies. The confining potential may be created either by the physical dimensions of the dot (using for example a small metallic grain [1] or material constriction [2, 3]), or by inducing an electrostatic potential at the surface of a semiconducting heterostructure [4]. The latter method is desirable for many applications, particularly quantum information devices, as it allows greater control over the geometry of the confinement potentials, may be realistically scaled up to incorporate more than one dot, and is also compatible with existing planar transistor fabrication techniques.

The conduction properties of the SET are defined by the geometry and potential of the dot region and the conducting leads. In contrast to conventional transistors, where conductance can be continuously reduced by increasing the magnitude of the gate bias, the SET conductance oscillates as a result of the addition or removal of single electrons. As the population of electrons is incrementally decreased by depleting the small semiconducting region, their number may be reduced down to a single electron before device pinch off. The use of the spin quantum number of a single confined electron has been demonstrated as a 'quantum bit' in such semiconductor systems [5]. Though widely studied in the GaAs material systems [6-8], electrostatically defined SETs in InSb quantum well-based structures are unreported. The recent demonstration of acceptably low leakage Schottky gates patterned onto InSb/AlInSb heterostructure material [9, 10] allows us to demonstrate an electrostatically defined SET in an InSb-based material. Of all the III-V semiconductors InSb offers the smallest electron effective mass, the highest mobility and the largest g-factor (-51). The large g-factor has important implications for potential spin-to-charge readouts of quantum bits [11] and also offers the possibility of localised qubit addressing [12].

**Material and Device Fabrication**
The InSb/AlInSb heterostructure material was grown by solid-source molecular beam epitaxy (MBE) on a semi-insulating GaAs substrate. The structure, which is illustrated in fig.1, consists of an accommodation layer, a 3µm $Al_xIn_{1-x}Sb$ buffer (x=0.15), a 20nm InSb quantum well, followed by a 50nm $Al_xIn_{1-x}Sb$ (x=0.20) cap with Te modulation δ-doping (~$1x10^{12}cm^{-2}$) located 5nm above the quantum well. This forms a type I heterostructure, providing confinement for both electrons and holes in the quantum well channel. Hall measurements for this material have determined the mobility to be 35,000$cm^2V^{-1}s^{-1}$ (RT)/55,000$cm^2V^{-1}s^{-1}$ (77K) with a carrier concentration of $4.6x10^{11}cm^{-2}$ (RT)/ $2.7x10^{11}cm^{-2}$ (77K). The SET confinement potential was created using a 'fork-gate' arrangement similar to that reported by Meirav *et al* [13] and indicated in fig. 1.

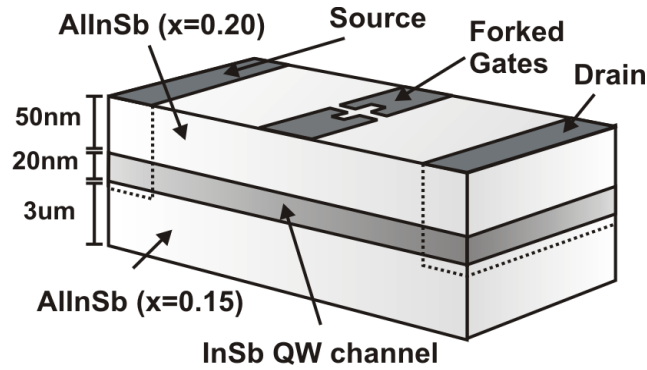

**Figure 1.** Schematic of an InSb/AlInSb quantum well heterostructure showing the layer structure and gate layout

The devices were fabricated using optical lithography to define the source and drain contacts using Ti/Au layers deposited by e-beam evaporation. E-beam lithography was used to define Ti/Au Schottky fork-gate structures onto mesas of width 3, 6 or 12µm. These gate contacts are approximately 750nm across and separated by 100nm at the constrictions at each end rising to a 300nm gap in the central (quantum dot) region. The devices were then isolated by wet chemical etching. The gates are fed in from either side and are air-bridged between the mesa edge and the feed metal to minimise current leakage due to material or surface conduction. Details of related InSb/AlInSb quantum well FETs with room temperature AC and DC performance can be found in reference [14], and details of Schottky barriers on this material can be found in references [10, 15]. The devices created by this process are intentionally depletion mode, and consequently are designed to operate under negative gate biases.

**Single Electron Charge Transport**

The addition energy spectrum of a single electron transistor can be accessed by sweeping the gate voltage so that the potential on the dot varies between pinch-off and some point at which the dot is no longer defined. These measurements were performed in a liquid helium bath cryostat at 4.5K in the presence of a small DC source-drain bias (100μV supplied from a HP4155B semiconductor parameter analyser). The gate voltage was swept between -0.29V and -0.25V, and the conductance was deduced from measurements of the drain current. At voltages below -0.29V the conductance peaks vanish, indicating that either the tunneling barriers are too large to allow further discernable charge transport or the dot is entirely depleted. The conductance of a typical SET device is plotted against the gate voltage in fig.2. Sharp periodic peaks are observed separated by regions of low conductance.

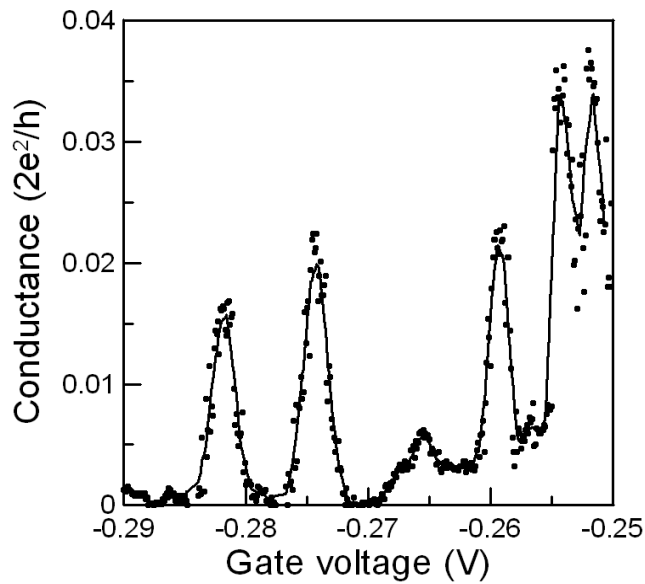

**Figure 2.** Conductance peaks observed in a typical InSb/AlInSb heterostructure quantum dot at 4.5K

The height of the peaks is significantly below the quantum limit of conductance for a single transverse mode, $2e^2/h$ [16], which suggests that the tunneling barriers leading to and from the dot are large. A consequence of this observation is that it is not possible to state with certainty that the last peak in fig. 2 represents the transition between zero and single electron occupancy.

Direct observation of the energy spectrum may also be provided by varying the drain voltage with a fixed gate voltage [8]. This allows the Fermi energy in the drain contact to pass through the energy levels in the dot increasing the number of possible current channels incrementally. This can be seen in the output current of another similar InSb/AlInSb SET in fig.3, where there is some finite region of minimal conductance about zero drain voltage of the order of a few millivolts, followed on either side by stepwise increments in current. This data is clearly asymmetrical, with larger positive than negative drain bias required to access the confined states. We attribute this to physical asymmetry in the gate structure, due to limitations in the lithography. The simplistic gate pattern employed in these devices (fork-gate) does not allow independent control of the tunneling barriers, and so a certain amount of asymmetry is expected. As a result the source and drain tunnel barrier transmission probabilities are likely to differ, giving an attenuated source-drain conductance and causing significant asymmetry in the bias conditions needed to align the confined states within the SET with those in the contacts.

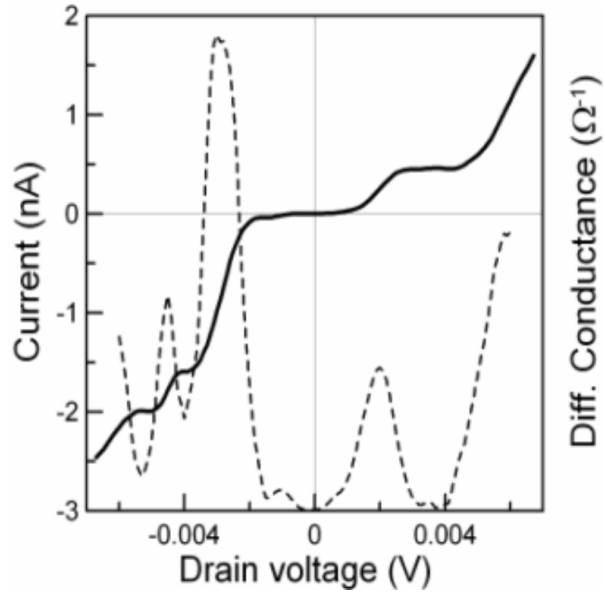

**Figure 3.** Output current through an InSb/AlInSb SET as a function of drain voltage, showing a very low conductance region at zero volts and stepwise increments in current for higher biases

**Single Electron Charging Model**
In considering the origin of the SET-like behaviour observed in figs. 2 and 3 we have adopted simple models for single particle confinement energy states in a circular dot and Coulomb blockade transport. The Coulomb blockade model describes how electron-electron repulsion results in an energy gap between the $N$ and $N+1$ charging states (where $N$ is the number of electrons on the dot) [17]. It is then predicted that in order to increase the electron population by one, the dot potential must be lowered by $U \sim e^2/C$, and that periodically in gate voltage there exist values of the dot potential where the $N$ and $N+1$ states are degenerate and single electrons can flow onto and off the dot freely (resulting in a current of single electrons). As well as this semi-classical approach it is also necessary to consider the quantised single particle states resulting from such strong confinement. In semiconducting materials with small effective masses, such as InSb, this consideration is important since the energy separation between single particle states, $\Delta E$, is inversely proportional to the square root of the effective mass, $m^*$ [18]. Much work has been done in describing the charging model of GaAs quantum dots in terms of the energy quantisation and Coulomb blockade, resulting in a shell-filling structure analogous to that of atomic physics [18, 19]. Generally for GaAs devices the dominant feature is Coulomb blockade, however the InSb-based dots studied here have both a larger $\Delta E$ and slightly smaller $e^2/C$ (due to the large dot size) so that, in principle, one can easily achieve a situation where $\Delta E \geq e^2/C > kT$, at temperatures up to a few tens of Kelvin. A two-dimensional calculation of the potential distribution in the quantum well of a 2DEG is shown in fig.4. This is generated for fork-gates with the same nominal dimensions as those observed from SEM inspection of our devices. The electrostatic approximation is based on that of Davies and Larkin [20]. The heavy dashed line illustrates the contour above which the 2DEG is depleted, and consequently defines the boundary of the quantum dot. This value was obtained by examining the potential required to deplete the ground state in the quantum well (2DEG) of the heterostructure according to a self-consistent Schrödinger-Poisson model [15]. At high biases the potential within the dot may, to a good approximation, be characterised in the form of a circular simple harmonic oscillator potential with easily calculable single particle energy states $E_n \approx (n_1+n_2+1)\hbar\omega$ (where $n_1$ and $n_2$ may be 0,1,2…), leading to a series of degenerate energy levels separated by $\hbar\omega$.

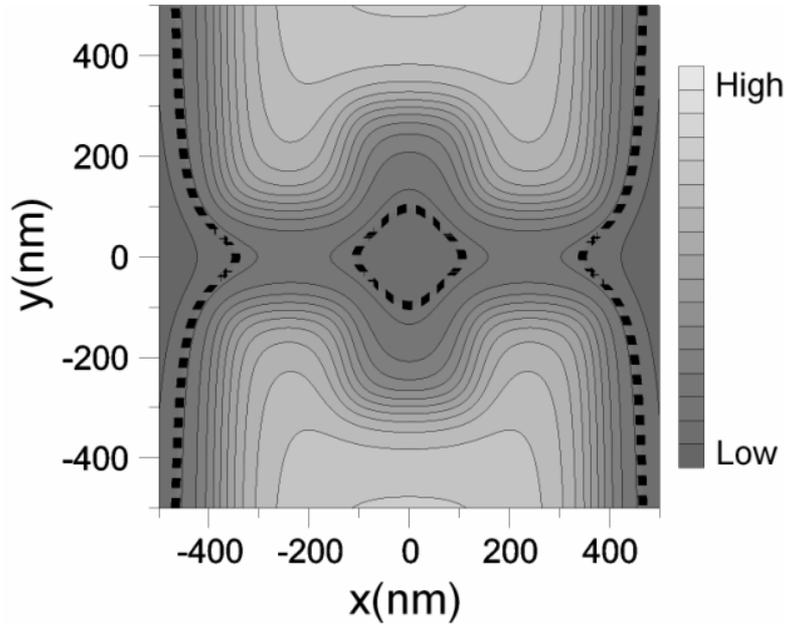

**Figure 4.** Approximate confinement potential generated by a typical fork-gated InSb/AlInSb SET with some finite negative bias

Deviation of the shape of the dot from a circular form will lead to a lifting of degeneracy in the confined states, and a more complicated energy spectrum which is not considered in this basic analysis. The circular confining potential was examined for a range of different gate biases, and the approximate Coulomb blockade charging energies and single particle states were calculated. It is found that the confining potential, defined by $V=\frac{1}{2}kr^2$, (where $r^2=(x^2+y^2)$ and $k=m^*\omega^2$) was approximately constant over the range of gate biases considered in this simulation, and the single particle state separation energy ($\Delta E$) was consistently of the order of a few meV. This is in good agreement with the charging energy spectrum (peak separation) observed in figure 2, and also the feature separation in drain voltage in figure 3 for a similar device. The capacitance of the SET is deduced from its radius (given by the dashed contour in fig. 4) as $C \propto \varepsilon r$ (where $\varepsilon$ is the permittivity of InSb), and from it the Coulomb blockade charging energy ($e^2/C$) is calculated. The capacitance of the ~100nm radius dot shown in fig.4 is approximately 0.1fF. The Coulomb blockade charging energy was seen to vary more strongly with gate voltage (as a result of the $1/r$ relationship) but remained similar to $\Delta E$, i.e. around a few meV, until the dot became extremely small, at which point no confined states remained. The feature separations in fig. 3 are consistent with a quantum dot energy spectrum with energy scales of the order of a few meV for both the Coulomb blockade and single particle states. The peak separation in fig.2 is suggestive of energy spacing smaller than ~7meV, since it must be considered that the potential dropped in the 2DEG is smaller than that applied to the gates due to the distribution of space charge in the heterostructure. It is therefore reasonable to conclude that the data in figs. 2 and 3 are consistent with SET charging models, where the Coulomb blockade and quantisation energy separation are both of the order of a few meV. In the absence of more substantial gate control it is difficult to elaborate on the precise shell structure of these devices, which would allow us to deduce the energy scales more accurately and individually.

**Conclusion**
We have demonstrated periodic conductance features in an electrostatically defined InSb single electron transistor, and estimated from simulation the quantisation energy to be of the same magnitude as the Coulomb blockade energy gap, with the potential to become larger if deeper confinement can be achieved, for instance by controlling the dot potential with a plunger gate [21] and smaller lithographically defined gates. The sharpness of these peaks at temperatures as high as 4.5K indicates that InSb may become an important material for quantum devices, where the low effective mass and high g-factor will provide significantly spaced energy levels for higher temperature operation and the potential for manipulation of electron spin.

**Acknowledgements**
This work was partially supported by the UK MOD, under output 4 of ES domain. J M S Orr acknowledges support from the UK EPSRC